\documentclass[12pt,a4paper,fleqn]{article}
\usepackage{amsmath,amssymb,amsfonts,amsthm,amsopn,graphicx}
\usepackage{tikz} 
\usetikzlibrary{fadings}
\usetikzlibrary{arrows}
\usepackage{hyperref}
\usepackage{mathrsfs} 
\usepackage{cite}

\usepackage{wrapfig}
\usepackage{float}
\usepackage{caption}
\usepackage{subcaption}

\usepackage[T1,T2A]{fontenc}
\usepackage[utf8]{inputenc}
\usepackage[main=english]{babel}

\allowdisplaybreaks[3]
\numberwithin{equation}{section} 

\newcommand{\ds}{\displaystyle}

\newcommand{\pt}{\mathbf{v}}

\newcommand{\kop}{\mathbf{k}}
\newcommand{\aop}{\mathbf{a}}
\newcommand{\bop}{\mathbf{b}}

\newcommand{\eop}{\mathbf{e}}

\newcommand{\Alg}{\mathcal{A}}

\newcommand{\hf}{{^1\!\!/\!_2}}

\def\GR{green!60!teal}
\def\BL{blue!90!}
\def\RD{red!90!}

\usepackage[body={15cm,22cm}]{geometry}

\begin{document}

\title{On spectral equations for an evolution operator of a $q$-oscillator lattice}
\author{Sergey Sergeev
    \thanks{
   Faculty of Science and Technology, 
   University of Canberra, Bruce ACT 2617, Australia
}
\thanks{ 
Department of Fundamental and Theoretical Physics,
         Research School of Physics and Engineering,
    Australian National University, Canberra, ACT 0200, Australia}
}

\date{}

\maketitle

\begin{abstract}
We propose a set of algebraic equations describing eigenvalues and eigenstates of a relativistic evolution operator for a two-dimensional $q$-oscillator Kagom\'e lattice. Evolution operator is constructed with the help of $q$-oscillator solution of the Tetrahedron Equation. We focus on the unitary regime of the evolution operator, so our results are related to 3d integrable systems of the quantum mechanics. Our conjecture is based on a two-dimensional lattice version of the coordinate Bethe-Ansatz.
\end{abstract}

\tableofcontents


\section{Introduction}

Solutions to the vertex-type Zamolodchikov's tetrahedron equation are known for many years
\cite{Zam:80,BS:82,BB:92,Kor:93,Hie:94,SMS:96,KV:91,BS:2006,S:2009,Kun:2024,BKMS:2025}. The main reason why all we were
interested in this subject is that the tetrahedron equation provides the Liouville's integrability of a corresponding lattice model of the statistical mechanics or of a corresponding lattice relativistic evolution model of the quantum mechanics.

This paper reports an attempt to derive a set of algebraic equations defining a spectrum of a spin lattice evolution operator. Such equations could stay as a multidimensional analogue of the Bethe-Ansatz equations for the spin chains.

In this introductory section we would like to discuss why we focus on evolution operator. When one considers a spin chain, the traditional quantum-mechanical object is its Hamiltonian. The evolution operator can also be introduced, but it has nothing common with the non-relativistics Hamiltonian. The evolution operator for a spin chain can be defined as follows:
\begin{equation}\label{ev2}
\begin{array}{l}
\ds \langle \boldsymbol{\sigma} | \boldsymbol{U} | \boldsymbol{\sigma}'\rangle \;=\;
\prod_{k\in\mathbb{Z}_M} \langle \sigma_{1,k+1}^{},\sigma_{2,k}^{} | R | \sigma_{1,k}',\sigma_{2,k}'\rangle\;=\\
\\
\ds 
\begin{tikzpicture}
\draw [-stealth, thin] (0,-1) -- (0,1); 
\draw [-stealth, thin] (-1.5,-1) -- (1.5,1);
\node [below] at (-1.5,-1) {\scriptsize $\sigma_{1,k}'$};
\node [below] at (0,-1) {\scriptsize $\sigma_{2,k}'$};
\node [above] at (0,1) {\scriptsize $\sigma_{2,k}$};
\node [above] at (1.5,1) {\scriptsize $\sigma_{1,k+1}$};
\draw [-stealth, thin] (0+3,-1) -- (0+3,1); 
\draw [-stealth, thin] (-1.5+3,-1) -- (1.5+3,1);
\node [below] at (-1.5+3,-1) {\scriptsize $\sigma_{1,k+1}'$};
\node [below] at (0+3,-1) {\scriptsize $\sigma_{2,k+1}'$};
\node [above] at (0+3,1) {\scriptsize $\sigma_{2,k+1}$};
\node [above] at (1.5+3,1) {\scriptsize $\sigma_{1,k+2}$};
\draw [-stealth, thin] (0+6,-1) -- (0+6,1); 
\draw [-stealth, thin] (-1.5+6,-1) -- (1.5+6,1);
\node [below] at (-1.5+6, -1) {\scriptsize $\sigma_{1,k+2}'$};
\node [below] at (0+6,-1) {\scriptsize $\sigma_{2,k+2}'$};
\node [above] at (0+6,1) {\scriptsize $\sigma_{2,k+2}$};
\node [above] at (1.5+6,1) {\scriptsize $\sigma_{1,k+3}$};
\node [left] at (-1.5,0) {$=\quad \cdots$};
\node [right] at (7.5,0) {$\cdots$};
\end{tikzpicture}
\end{array}
\end{equation}
Here the vertices are the traditional graphical representations of the matrix elements of the $R$-matrices of the Yang-Baxter equation. Evolution operators can also be used in statistical mechanics since 
\begin{equation}
Z\;=\;\textrm{Trace}\; \boldsymbol{U}^M
\end{equation}
is the partition function for the usual $M\times M$ square lattice in the regime of positive Boltzmann weights. However, the spirit of quantum mechanics implies a different physical regime, namely the regime of unitary $R$ and $\boldsymbol{U}$. The locality of the evolution operator (\ref{ev2}) corresponds to the relativistic principle of a causality.

The spin chain relativistic evolution operator (\ref{ev2}) evidently commutes with the usual auxiliary transfer-matrices and therefore the spectral equations for $\boldsymbol{U}$ evidently coincide with the usual Bethe-Ansatz equations for the length $2M$ spin chain \cite{Bax-arc}.

Local non–relativistic Hamiltonians for three-dimensional models are not known, but the idea of the relativistic evolution operators can be straightforwardly extended from the spin chains to the spin lattices as follows:
\begin{equation}\label{ev3}
\begin{array}{l}
\ds \langle \boldsymbol{\sigma} | \boldsymbol{U} |\boldsymbol{\sigma}'\rangle = 
\prod_{\pt\in\mathbb{Z}_M^2} \langle \sigma_{1,\pt+\eop_1}^{},\sigma_{2,\pt}^{},\sigma_{3,\pt+\eop_3}| R |\sigma_{1,\pt}',\sigma_{2,\pt}',\sigma_{3,\pt}'\rangle\;=\\
\\
\ds
\begin{tikzpicture}
\draw [-stealth, blue, thin] (0+1.5,-1+1) -- (0+1.5,1+1);
\draw [-stealth, blue, thin] (-2+1.5,-1+1) -- (2+1.5,1+1);
\draw [-stealth, blue, thin] (-1+1.5,-1.5+1) -- (1+1.5,1.5+1);
\draw [-stealth, blue, thin] (0+1.5+4,-1+1) -- (0+1.5+4,1+1);
\draw [-stealth, blue, thin] (-2+1.5+4,-1+1) -- (2+1.5+4,1+1);
\draw [-stealth, blue, thin] (-1+1.5+4,-1.5+1) -- (1+1.5+4,1.5+1);
\node [below] at (0+1.5+0.5,-1+1) {\textcolor{blue}{\scriptsize $\sigma_{2,\pt+\eop_3}'$}};
\node [below] at (0+1.5+4+0.5,-1+1) {\textcolor{blue}{\scriptsize $\sigma_{2,\pt+\eop_{13}}'$}};
\draw [-stealth, thick] (0,-1) -- (0,1);
\draw [-stealth, thick] (-2,-1) -- (2,1);
\draw [-stealth, thick] (-1,-1.5) -- (1,1.5);
\node [below] at (-2,-1) {\scriptsize $\sigma_{1,\pt}'$};
\node [below] at (-1,-1.5) {\scriptsize $\sigma_{3,\pt}'$};
\node [below] at (0,-1) {\scriptsize $\sigma_{2,\pt}'$};
\node [right] at (2,1) {\scriptsize $\sigma_{1,\pt+\eop_1}$};
\node [above] at (1,1.5) {\scriptsize $\sigma_{3,\pt+\eop_3}$};
\node [above] at (0,1) {\scriptsize $\sigma_{2,\pt}$};
\draw [-stealth, thick] (0+4,-1) -- (0+4,1);
\draw [-stealth, thick] (-2+4,-1) -- (2+4,1);
\draw [-stealth, thick] (-1+4,-1.5) -- (1+4,1.5);
\node [below] at (-2+4,-1) {\scriptsize $\sigma_{1,\pt+\eop_1}'$};
\node [below] at (-1+4,-1.5) {\scriptsize $\sigma_{3,\pt+\eop_1}'$};
\node [below] at (0+4,-1) {\scriptsize $\sigma_{2,\pt+\eop_1}'$};
\node [right] at (2+4,1) {\scriptsize $\sigma_{1,\pt+2\eop_1}$};
\node [above] at (1+4-0.2,1.5) {\scriptsize $\sigma_{3,\pt+\eop_{13}}$};
\node [above] at (0+4,1) {\scriptsize $\sigma_{2,\pt+\eop_1}$};
\node [left] at (-2,0) {$=\quad \cdots$};
\node [right] at (7.5,0) {$\cdots$};
\end{tikzpicture}
\end{array}
\end{equation}
The vertices here are the graphical representation of the $R$-matrices of the vertex-type Zamolodchikov's Tetrahedron Equation. This picture of the evolution system in $2+1$ dimensional space-time was proposed in e.g. \cite{K:95}. 
Similarly to the two-dimensional case,
\begin{equation}
Z\;=\;\textrm{Trace}\; \boldsymbol{U}^M
\end{equation}
could be the partition function for $M\times M\times M$ cubic lattice. However, we focus on the relativistic field theory with unitary $R$ and $\boldsymbol{U}$ rather than on statistical mechanics with the real matrix elements of $R$.

The "space-type surface" now is the auxiliary lattice equipped by the set of "spin indices" $\sigma_{1,\pt},\sigma_{2,\pt},\sigma_{3,\pt}$, where
$\pt=(k,\ell)=k\eop_1+\ell\eop_3$,  and $k,\ell\in\mathbb{Z}_M$. Wel call it "spin lattice".

It is important to note, the auxiliary evolution lattice is not a square lattice. It is called after R. J. Baxter the ``Kagom\'e lattice'' \cite{Bax-book}, see picture (\ref{kar1}) below. It means \`a priori, one could not expect that the Nested Bethe-Ansatz equations will appear in the framework of the spectral problem for the evolution operator since the Nested Bethe-Ansatz equations are essentially the subject of the square auxiliary lattice (see e.g. \cite{S:2006} for the Nested Bethe-Ansatz structure in three-dimensional models). However, the evolution operator commutes with a Kagom\'e layer-to-layer transfer matrix, and therefore the evolution is integrable in Liouville's sense.
\\

We consider in this paper the evolution operator (\ref{ev3}) with the $q$-oscillator $R$-matrix of the Tetrahedron Equation in the Fock space representation \cite{BS:2006}. The advantage of the Fock space representation is that the total Fock vacuum is the ground state for the evolution operator, and also the rather complicated form of the matrix elements of $R$-matrix is not required for our considerations, the evolution operator is uniquely defined by its adjoint action on local creation and annihilation operators.
\\

A method we use for a construction of eigenstates is a form of coordinate Bethe-Ansatz. The method works in general, however it has essentially more complicated structure in comparison with the spin-chain Bethe-Ansatz, see e.g. Appendix.
We do not have a general formula for eigenstates. We constructed explicitly the eigenstates for relatively small occupation numbers, and observe some common and rather general properties of the spectral equations. It allows us to make a conjecture about a general form of the spectral equations. This conjecture is presented in this paper.
As expected, the spectral equations obtained have nothing common with the Nested Bethe-Ansatz equations.
\\

This paper is organised as follows. In Section \ref{section1} we give a more detailed formulation of the $q$-oscillator evolution system on the Kagom\'e lattice. Then, in Section \ref{section2} we discuss briefly the idea of the coordinate Bethe-Ansatz (more technical details are given in Appendix \ref{AA}). In Section \ref{section3} we formulate our hypothesis about the structure of the spectral equations for the evolution model. Finally, we discuss our results in Section \ref{conclusion}.

\section{Formulation of the Model}\label{section1}

Let the set of the generators 
\begin{equation}
\Alg\;\;=\;\; [1,\aop^+,\aop^-,\kop,\kop']
\end{equation}
satisfy the $q$-oscillator algebra relations:
\begin{equation}
\aop^+\aop^-=1+q^{-1}\kop\kop'\;,\;\;
\aop^-\aop^+=1+q\kop\kop'\;,\;\;
\kop^{\#}\aop^{\pm}\;=\;q^{\pm 1}\aop^{\pm}\kop^{\#}\;.
\end{equation}
The Fock space representation is defined by 
\begin{equation}
\aop^{-} |0\rangle\;=\;0\;,\quad 
(\aop^{+})^n | 0 \rangle \;\sim\; |n\rangle\;,\quad
\kop |n\rangle \;=\; -\kop'|n\rangle \;=\; |n\rangle q^{\hf+n}\;.
\end{equation}
The representation is unitary if $(\aop^{-})^\dagger = \aop^{+}$ and $0<q<1$.

Before we formulate the evolution model, we need to mention the auxiliary tetrahedron equation.
The three-dimensional auxiliary $L_{\alpha,\beta}(\Alg)$-operator can be defined in Baxter's notations by
\begin{equation}
\begin{tikzpicture}[scale=0.75]
\draw [-stealth, thin] (1-6,3) -- (1-6,5);
\draw [-stealth, thin] (0-6,4) -- (2-6,4);
\node [below] at (1-6,3) {\scriptsize $\alpha$};
\node [left] at (0-6,4) {\scriptsize $\beta$};
\node [right] at (2.1-6,4.05) {$=1$};
\draw [-stealth, ultra thick] (1-6,0) -- (1-6,2);
\draw [-stealth, ultra thick] (0-6,1) -- (2-6,1);
\node [below] at (1-6,0) {\scriptsize $\alpha$};
\node [left] at (0-6,1) {\scriptsize $\beta$};
\node [right] at (2.1-6,1.05) {$=1$};
\draw [-stealth,ultra thick] (1,3) -- (1,5);
\node [below] at (1,3) {\scriptsize $\alpha$};
\draw [-stealth, thin] (0,4) -- (2,4);
\node [left] at (0,4) {\scriptsize $\beta$};
\node [right] at (2.1,4.05) {$=\kop$};
\draw [thin] (1,0) -- (1,1); 
\draw [-stealth, thin] (1,1) -- (2,1);
\draw [ultra thick] (0,1) -- (1,1); 
\draw [-stealth, ultra thick] (1,1) -- (1,2);
\node [right] at (2.1,1.05) {$=\aop^{-}$};
\node [left] at (0,1) {\scriptsize $\beta$};
\node [below]  at (1,0) {\scriptsize $\alpha$}; 
\draw [thin] (4+1,4) -- (5+1,4);
\draw [-stealth, thin] (5+1,4) -- (5+1,5);
\draw [ultra thick] (5+1,3) -- (5+1,4);
\draw [-stealth, ultra thick] (5+1,4) -- (6+1,4);
\node [below] at (5+1,3) {\scriptsize $\alpha$};
\node [left] at (4+1,4) {\scriptsize $\beta$};
\node [right] at (6.1+1,4.05) {$=\aop^{+}$};
\draw [-stealth, ultra thick] (4+1,1) -- (6+1,1);
\draw [-stealth, thin] (5+1,0) -- (5+1,2); 
\node [right] at (6.1+1,1.05) {$=\kop'$};
\node [left] at (4+1,1) {\scriptsize $\beta$};
\node [below]  at (5+1,0) {\scriptsize $\alpha$}; 
\end{tikzpicture}
\end{equation}
or, in the matrix notations, by
\begin{equation}
L_{\alpha,\beta}(\Alg)\;=\;\left(\begin{array}{cccc}
1 & 0 & 0 & 0  \\
0 & \kop & \aop^{+} & 0 \\
0 & \aop^{-} & \kop' & 0  \\
0 & 0 & 0 & 1
\end{array}\right)\;.
\end{equation}
The auxiliary Tetrahedron Equation, or the local Yang-Baxter equation, is 
\begin{equation}\label{LYBE}
L_{\alpha,\beta}(\Alg_1) L_{\alpha,\gamma}(\Alg_2) L_{\beta,\gamma}(\Alg_3) R_{123}\;=\;
R_{123} L_{\beta,\gamma}(\Alg_3) L_{\alpha,\gamma}(\Alg_2) L_{\alpha,\beta}(\Alg_1)\;,
\end{equation}
where $\Alg_1=\Alg\otimes 1\otimes 1$, $\Alg_2=1\otimes \Alg\otimes 1$, etc., are the local $q$-oscillator algebras, and the indices in $R_{123}$ refer to the components in the tensor product.
In the two-dimensional graphical representation it is
\begin{equation}
\begin{tikzpicture}
\draw [-stealth, thick, \BL] (0,-1) -- (0,3);
\draw [-stealth, thick, \RD] (3,0) -- (-1,0);
\draw [-stealth, thick, \GR] (2.5,-0.5) -- (-0.5,2.5);
\node [above] at (-0.6,2.5) {\scriptsize $\beta$};
\node [above] at (0,3) {\scriptsize $\gamma$};
\node [right] at (3,0) {\scriptsize $\alpha$};
\node [above] at (2.2,0) {\scriptsize $\Alg_1$};
\node [right] at (0.1,2) {\scriptsize $\Alg_3$};
\node [right] at (0,0.25) {\scriptsize $\Alg_2$};
\node [right] at (3,1) {$\times\, R_{123}\quad\quad=$};
\end{tikzpicture}
\qquad
\begin{tikzpicture}
\node [left] at (-3,-1) {$R_{123}\, \times$};
\draw [-stealth, thick, \BL] (0,-3) -- (0,1);
\draw [-stealth, thick, \RD] (1,0) -- (-3,0);
\draw [-stealth, thick, \GR] (0.5,-2.5) -- (-2.5,0.5);
\node [above] at (-2.5,0.5) {\scriptsize $\beta$};
\node [above] at (0,1) {\scriptsize $\gamma$};
\node [right] at (1,0) {\scriptsize $\alpha$};
\node [right] at (0,0.25) {\scriptsize $\Alg_2$};
\node [above] at (-1.8,0) {\scriptsize $\Alg_1$};
\node [right] at (0,-1.9) {\scriptsize $\Alg_3$};
\end{tikzpicture}
\end{equation}
In the three-dimensional graphical representation it could be the equivalence of two tetrahedra, 
the three-dimensional representation appeared in picture (\ref{ev3}), but in what follows 
the three-dimensional graphics is useless.

The evolution operator appears when one extends the definition of the tetrahedral map to the whole evolution lattice,
\begin{equation}
\begin{tikzpicture}
\draw [-stealth, thick, \BL] (0,-1) -- (0,3);
\draw [-stealth, thick, \RD] (3,0) -- (-1,0);
\draw [-stealth, thick, \GR] (2.5,-0.5) -- (-0.5,2.5);
\node [above] at (-0.6,2.5) {\scriptsize $\beta_{j+1}$};
\node [above] at (0,3) {\scriptsize $\gamma_k$};
\node [right] at (3,0) {\scriptsize $\alpha_\ell$};
\node [above] at (2.4,0) {\scriptsize $\Alg_{1,\pt+\eop_1}$};
\node [right] at (0.1,2) {\scriptsize $\Alg_{3,\pt+\eop_3}$};
\node [right] at (0,0.25) {\scriptsize $\Alg_{2,\pt}$};
\node [right] at (3,1) {$\times\, \boldsymbol{U}\quad\quad=$};
\end{tikzpicture}
\qquad
\begin{tikzpicture}
\node [left] at (-3,-1) {$\boldsymbol{U}\, \times$};
\draw [-stealth, thick, \BL] (0,-3) -- (0,1);
\draw [-stealth, thick, \RD] (1,0) -- (-3,0);
\draw [-stealth, thick, \GR] (0.5,-2.5) -- (-2.5,0.5);
\node [above] at (-2.5,0.5) {\scriptsize $\beta_j$};
\node [above] at (0,1) {\scriptsize $\gamma_k$};
\node [right] at (1,0) {\scriptsize $\alpha_\ell$};
\node [right] at (0,0.25) {\scriptsize $\Alg_{2,\pt}$};
\node [above] at (-1.8,0) {\scriptsize $\Alg_{1,\pt}$};
\node [right] at (0,-1.9) {\scriptsize $\Alg_{3,\pt}$};
\end{tikzpicture}
\end{equation}
As a formula, this is 
\begin{equation}\label{TU}
\begin{array}{l}
\ds
\biggl(\cdots L_{\alpha_\ell,\beta_{j+1}}(\Alg_{1,\pt+\eop_1}) L_{\alpha_\ell,\gamma_k}(\Alg_{2,\pt}) 
L_{\beta_{j+1},\gamma_k}(\Alg_{3,\pt+\eop_3}) \cdots \biggr)\, \times\, \boldsymbol{U} \;=\;\\
\\
\ds \qquad \qquad =\;
\boldsymbol{U}\, \times \, \biggl(\cdots 
L_{\beta_{j},\gamma_k}(\Alg_{3,\pt}) L_{\alpha_\ell,\gamma_k}(\Alg_{2,\pt}) 
L_{\alpha_\ell,\beta_j}(\Alg_{1,\pt})\cdots\biggr)\;.
\end{array}
\end{equation}
Here $\pt$  is the point on two-dimensional lattice,
\begin{equation}
\pt \;=\; (k,\ell) \;=\; k\eop_1 + \ell\eop_3
\end{equation}
A bigger fragment of the auxiliary evolution lattice is the following:
\begin{equation}\label{kar1}
\begin{tikzpicture}[scale=0.6]
\draw [-stealth, thick, \RD] (10,0) -- (-4,0);
\draw [-stealth, thick, \RD] (10,6) -- (-4,6);
\node [left] at (-4,0) {\scriptsize $\alpha_\ell$};
\node [left] at (-4,6) {\scriptsize $\alpha_{\ell+1}$};
\draw [-stealth, thick, \BL] (0,-4) -- (0,10);
\draw [-stealth, thick, \BL] (6,-4) -- (6,10);
\node [above] at (0,10) {\scriptsize $\gamma_k$};
\node [above] at (6,10) {\scriptsize $\gamma_{k+1}$};
\draw [-stealth, thick, \GR] (1,-4) -- (-4,1);
\draw [-stealth, thick, \GR] (7,-4) -- (-4,7);
\draw [-stealth, thick, \GR] (10,-1) -- (-1,10);
\node [above] at (-4,1) {\scriptsize $\beta_j$};
\node [above] at (-4,7) {\scriptsize $\beta_{j+1}$};
\node [above] at (-1,10) {\scriptsize $\beta_{j+2}$};
\node [right] at (0,0.3) {\scriptsize $\mathcal{A}_{2,(k,\ell)}$};
\node [right] at (0,6.3) {\scriptsize $\mathcal{A}_{2,(k,\ell+1)}$};
\node [right] at (6,0.3) {\scriptsize $\mathcal{A}_{2,(k+1,\ell)}$};
\node [right] at (6,6.3) {\scriptsize $\mathcal{A}_{2,(k+1,\ell+1)}$};
\node [right] at (0,-3) {\scriptsize $\mathcal{A}_{3,(k,i)}$};
\node [right] at (0,3) {\scriptsize $\mathcal{A}_{3,(k,\ell+1)}$};
\node [right] at (6,-3) {\scriptsize $\mathcal{A}_{3,(k+1,\ell)}$};
\node [right] at (6,3) {\scriptsize $\mathcal{A}_{3,(k+1,\ell+1)}$};
\node [right] at (-3.1,0.3) {\scriptsize $\mathcal{A}_{1,(k,i)}$};
\node [right] at (-3.1+6,0.3) {\scriptsize $\mathcal{A}_{1,(k+1,\ell)}$};
\node [right] at (-3.1,0.3+6) {\scriptsize $\mathcal{A}_{1,(k,\ell+1)}$};
\end{tikzpicture}
\end{equation}
This lattice is called the Kagom\'e lattice.
Blue and red lines form the simple square lattice. The $q$-oscillator algebras $\mathcal{A}_{2,(k,\ell)}$ inhabit its vertices. The green lines have the slope $3\pi/4$, the intersections of the green lines with the blue and red lines inhabit the $q$-oscillator algebras $\mathcal{A}_{1,(k,i)}$ and $\mathcal{A}_{3,(k,i)}$ as it is shown in (\ref{kar1}). Numeration is: $k$ grows to the right and $\ell$ grows up. As to the numeration of the green lines, one may assume $j=k+\ell$. Let $M$ be the size of the lattice in both directions, and the periodical boundary conditions assumed:
\begin{equation}
i,j,k\;\in\;\mathbb{Z}_M\;.
\end{equation}
The evolution operator $\boldsymbol{U}$ is Liouville-integrable by construction. To deduce it, one could consider the trace of both left- and right-hand sides of (\ref{TU}) with respect to auxiliary red $\alpha_\ell$, blue $\gamma_k$ and green $\beta_j$ spaces with the quasi-periodical boundary conditions. The product of $L$-operators produces an auxiliary transfer matrix commuting with $\boldsymbol{U}$ and generating a complete set of integrals of motion \cite{BS:2006,S:2009,S:2006,Kuniba:2025}.
\\

The advantage of the $q$-oscillator model is that the local Yang-Baxter equation (\ref{LYBE}) defines straightforwardly and ``almost'' explicitly the tetrahedral map of the algebra of observables. The definition of the map can be extended naturally to the whole evolution operator. It can be written as 
\begin{equation}\label{Umap}
\left\{
\begin{array}{l}
\ds \boldsymbol{U} \; \kop_{2,\pt}^{} \aop_{1,\pt}^{+} \; \boldsymbol{U}^{-1}\;=\; 
\kop_{3,\pt+\eop_3}^{} \aop_{1,\pt+\eop_1}^{+} + \kop_{1,\pt+\eop_1}^{} \aop_{2,\pt}^{+} \aop_{3,\pt+\eop_3}^{-}\;,\\
\\
\ds \boldsymbol{U} \; \aop_{2,\pt}^{+} \; \boldsymbol{U}^{-1}\;=\; 
\aop_{1,\pt+\eop_1}^{+}\aop_{3,\pt+\eop_3}^{+} + \kop_{1,\pt+\eop_1}^{}\kop_{3,\pt+\eop_3}' \aop_{2,\pt}^{+}\;,\\
\\
\ds \boldsymbol{U} \; \kop_{2,\pt}' \aop_{3,\pt}^{+} \; \boldsymbol{U}^{-1}\;=\; 
\kop_{1,\pt+\eop_1}' \aop_{3,\pt+\eop_3}^{+} + \kop_{3,\pt+\eop_3}' \aop_{2,\pt}^{+} \aop_{1,\pt+\eop_1}^{-}\;.
\end{array}
\right.
\end{equation}
where the simple boundary conditions imply
\begin{equation}
\mathcal{A}_{j,\pt+M\eop_1}\;=\;
\mathcal{A}_{j,\pt+M\eop_3}\;=\;\mathcal{A}_{j,\pt}\;.
\end{equation}
Matrix elements of the evolution operator are given by the top relation in (\ref{ev3}), but since we have relations (\ref{Umap}), the matrix elements are in fact not required. The main working formula is now the very simple formula (\ref{Umap}).

\section{The coordinate Bethe-Ansatz method}\label{section2}

The ground state for the coordinate Bethe-Ansatz method is the total Fock vacuum,
\begin{equation}
\boldsymbol{U} | 0\rangle \;=\; | 0 \rangle\;.
\end{equation}
The next evident step is to construct a ``one-particle state''. It is simple. Let 
\begin{equation}\label{one-A}
A^+_\pt(u) \;=\; \aop_{2,\pt}^{+} + \sum_{k=1}^M g_\pt^{(k)} u^{-k} \aop_{1,\pt+k\eop_1}^+\aop_{3,\pt+k\eop_3}^+\;.
\end{equation}
This operator gives the elementary eigenstate of the evolution operator,
\begin{equation}\label{one-state}
\boldsymbol{U} A^+_\pt(u) |0\rangle \;=\; A^+_\pt(u) | 0 \rangle \,u\;,
\end{equation}
if the cyclicity condition is satisfied,
\begin{equation}\label{one-sp}
\frac{1+qu}{1-q^2} = g_\pt^{(1)} = g_{\pt}^{(2)} = \cdots  = g_{\pt}^{(M)} = \frac{q+u}{1-q^2} u^M\;,
\end{equation}
so that the easiest spectral equation is
\begin{equation}\label{sp-1}
\frac{q+u}{1+qu}\, u^M \;=\; 1\;.
\end{equation}
This can be easily verified. Using (\ref{Umap}), one obtains straightforwardly
\begin{equation}\label{U1}
\left\{
\begin{array}{l}
\ds \boldsymbol{U}\, \aop_{1,\pt+k\eop_1}^+\aop_{3,\pt+k\eop_3}^+|0\rangle = \aop_{1,\pt+(k+1)\eop_1}^+\aop_{3,\pt+(k+1)\eop_3}^+|0\rangle,\quad k\neq M\;,\\
\\
\ds \boldsymbol{U}\, \aop_{1,\pt}^+ \aop_{2,\pt}^+ |0\rangle  =
\left( (1-q^2)\aop_{2,\pt}^{+}+q\aop_{1,\pt+\eop_1}^+\aop_{3,\pt+\eop_3}^+\right) |0\rangle \;,\\
\\
\ds \boldsymbol{U}\, \aop_{2,\pt}^{+} |0\rangle = \left( -q\aop_{2,\pt}^{+}+\aop_{1,\pt+\eop_1}^+\aop_{3,\pt+\eop_3}^+\right) | 0\rangle \;.
\end{array}\right.
\end{equation}
Substituting (\ref{U1}) into (\ref{one-state}), one gets (\ref{one-sp}), and then one comes to (\ref{sp-1}).
\\

The one-particle state can be interpreted as follows. A boson of type $2$ (a particle, or an impurity) sits at the vertex $\pt$. It decays into a pair of two correlated bosons of types $1$ and $3$ (photons). Turning around the torus, the photons recombine back into the impurity.
This interpretation arise for the elementary excitations over the Fock vacuum. It remains unclear how excitations with big occupation numbers can be interpreted.
\\

In this paper we discuss the eigenstates of the type\footnote{Such set is not complete. E.g. I did not consider 
extra photons, and I did not consider the states like $A_{0,0}^+A_{n_1,m_1}^+\oplus A_{n_1,0}^+A_{0,m_1}^{+}$.}
\begin{equation}\label{BA}
|\Psi_N \rangle \;=\; \sum_{\sigma } C_{\sigma} \; :\; A^+_{\pt_1}(u_{\sigma_1}) 
A^+_{\pt_2}(u_{\sigma_2}) \cdots 
A^+_{\pt_N}(u_{\sigma_N})\;:\;|0\rangle\;,
\end{equation} 
where $\sigma$ is the set of $N!$ permutations of the indices $1,2,\cdots,N$,
\begin{equation}
\sigma\;:\;\;1,2,\cdots,N\;\to \sigma_1,\sigma_2,\cdots,\sigma_N\;,
\end{equation} 
and 
\begin{equation}\label{BA-eigen}
\boldsymbol{U}\; |\Psi_N\rangle \;=\; |\Psi_N\rangle\, u_1u_2\cdots u_N\;.
\end{equation}
The double dots in (\ref{BA}) mean that after expansion of the brackets in the product of $A_{\pt}^{+}$ the expansion coefficients require some corrections. For instance, formula (\ref{one-A}) corresponds to a ``general position'', while after the expansion of brackets there appear a set of ``special'' points when the creation operators from different $A_{\pt}^{+}$ join together at one point. In general, the problem of explicit construction of (\ref{BA}) appears essentially more complicated than the usual spin-chain Bethe-Ansatz. For example, the coefficients $C_\sigma$ essentially depend on a two-dimensional geometry of the set of $\pt_1,\pt_2,\cdots,\pt_N$, i.e. on how this set is distributed across the lattice.  Also, the coefficients $C_\sigma$ do not factorise into two-particles factors as it happens in the spin-chain case, and moreover, the coefficients $C_\sigma$ are not rational functions of $u_1,u_2,\cdots,u_N$.
\\

We have constructed explicitly several particular examples of (\ref{BA}) explicitly for relatively small occupation numbers $N\leq 4$ and for various distributions of $\pt_1,\pt_2,\pt_3,\pt_4$. The general case is roughly described in the Appendix \ref{AA}.
However, all the particular cases constructed allows one to formulate a hypothesis about a general form of the spectral equations for the Ansatz (\ref{BA}). The hypothesis is formulated in the next section.

\section{General form of the spectral equaitons}\label{section3}

Initial variables in (\ref{BA}) and in (\ref{BA-eigen}) are $u_j$. 
However, it is convenient to introduce variables $x_j$ and $X_j$ related to $u_j$ as follows:
\begin{equation}
u_j\;=\;\frac{x_j-q}{1-qx_j}\;,\quad  x_j \;=\; \frac{q+u_j}{1+qu_j}\;,\quad X_j\;=\; u_j^M \frac{q+u_j}{1+qu_j}\;.
\end{equation}
Let us define also 
\begin{equation}
S_{i,j}\;=\;\frac{q+qu_i+q^2u_i+qu_iu_j}{q(u_i-u_j)}\;=\;\frac{q^{-1}x_i-qx_j}{x_i-x_j}\;.
\end{equation}
Next, let us define the following set of symmetric combinations:
\begin{equation}\label{mysym}
\left\{\quad
\begin{array}{l}
\ds \mathcal{F}_1\;\stackrel{def}{=}\; 
\sum_{i=1}^N X_i \prod_{j\neq i} S_{i,j}\\
\\
\ds \mathcal{F}_2\;\stackrel{def}{=}\; 
\sum_{i_1<i_2} X_{i_1}X_{i_2} \prod_{j\neq i_1,i_2} S_{i_1,j} S_{i_2,j}\\
\vdots\\
\ds \mathcal{F}_n\;\stackrel{def}{=}\;
\sum_{i_1<\cdots <i_n} X_{i_1}\cdots X_{i_n} \prod_{j\neq i_1,\cdots,i_n} S_{i_1,j} \cdots S_{i_n,j}\\
\vdots \\
\ds \mathcal{F}_{N-1}\;\stackrel{def}{=}\;
X_1\cdots X_N \sum_{i=1}^N X_i^{-1} \prod_{j\neq i} S_{j,i}\\
\\
\ds \mathcal{F}_N\;\stackrel{def}{=}\;X_1\cdots X_N\;,\quad \mathcal{F}_0\;\stackrel{def}{=}\;1\;.
\end{array}\right.
\end{equation}
Then the spectral equations are 
\begin{equation}\label{mysp}
\mathcal{F}_{n}\;=\;P_{n,N}(q)\;,\quad n=1,\cdots,N,
\end{equation}
where $P_{n,N}(q)$ are some symmetric Laurent polynomials of $q$. The form of these polynomials depends on the geometry of $\pt_1,\pt_2,\cdots, \pt_N$. The best way to classify the polynomials $P_{n,N}(q)$ is to introduce their generation function
\begin{equation}
\mathcal{P}_N(z;q)\;=\;\sum_{n=0}^N z^n P_{n,N}(q)\;.
\end{equation}
Three particular cases of the polynomials $\mathcal{P}_N(z;q)$ are to be mentioned here:
\begin{itemize}
\item When $N$ particles are situated along a single vertical or horizontal line, for instance 
\begin{equation}
\pt_1=(k_1,0),\;\;
\pt_2=(k_2,0),\;\;
\cdots,\;\;
\pt_N=(k_N,0)\;,
\end{equation}
then 
\begin{equation}
P_{n,N}(q) \;=\; {N\choose n}\;,\quad 
\mathcal{P}_N(z;q)\;=\;(1+z)^N\;.
\end{equation}
\item
When all $N$ particles sit at the same point,  this is the state $\ds \prod_{j=1}^N A_{\pt}^+(u_j) \, |0\rangle$, or all $N$ particles are distributed on the plane randomly, then 
\begin{equation}
P_{n,N}(q)\;=\;q^{-n(N-n)}{N\choose n}_{q^2}\;,\quad
\mathcal{P}_N(z;q)\;=\;(-q^{1-N}z;q^2)_N\;.
\end{equation}
In both cases equations (\ref{mysp}) can be simplified, their solution set has two outmost branches. One branch is 
\begin{equation}\label{X1}
X_1=X_2=\cdots=X_N=1\;.
\end{equation}
This branch correspond to $N$ randomly distributed non-interacting perticles.
The second branch is 
\begin{equation}\label{XXZ}
X_i\;=\;\prod_{j\neq i} \frac{S_{j,i}}{S_{i,j}}\;,
\end{equation}
it corresponds to $N$ particles sitting at the same point. It is remarkable to note, equations (\ref{XXZ}) coincide with the XXZ spin $-\hf$ chain Bethe-Ansatz equations.
\item Finally, the main part of our hypothesis about the $q$-oscillator lattice spectral equations: if the $A_{\pt}^+$-particles are 
situated at the vertices of a sub-lattice,  $N=K\times L$, where $K\geq L$ and
\begin{equation}
\{\pt\}\;=\;\{(n_k \eop_1+m_\ell\eop_3)\}\;,\quad k=1,\cdots,K,\;\;\ell =1,\cdots,L
\end{equation}
then 
\begin{equation}\label{myconj}
\mathcal{P}_{K\times L}(z;q)\;=\;(-q^{1-L}z;q^2)_L^K\;,\;\; K\geq L\;.
\end{equation}
\end{itemize}

Concluding this section, one could mention two general properties of equations (\ref{mysp}). The first property is that the map
\begin{equation}
X_i\;\to\; X_i^{-1}\prod_{j\neq i} \frac{S_{j,i}}{S_{i,j}}
\end{equation}
is the symmetry of the system for all known cases. The second property is that 
\begin{equation}
P_{n,N}(q) \;\to \; {N\choose n}\;\;\textrm{when}\;\; q\to 1
\end{equation}
for all known cases as well. Also note that $P_{N,N}(q)=1$ in all cases, so that we always have $X_1X_2\cdots X_N=1$.

\section{Discussion}\label{conclusion}

We present in this paper the hypothesis -- collection of formulas (\ref{mysym},\ref{mysp},\ref{myconj}) -- for the spectral equations defining the eigenvalues of the evolution operator of the $q$-oscillator Kagom\'e lattice.

The result presented requires several critical notes. Out Ansatz (\ref{BA}) corresponds to a part of possible eigenstates, it does not take into account ``extra free photons'' and it does not take into account more general combinatorics for the states distributed along the auxiliary Kagom\'e lattice. However, Ansatz (\ref{BA}) corresponds to the most interesting subspace of the whole Hilbert space corresponding to regular patterns of $\pt_1,\pt_2,\cdots,\pt_N$.
Also, our approach is not as effective as the standard Bethe-Ansatz for the spin chains. The reader may see Appendix \ref{AA} as a demonstration of how clumsy is our approach. Our hypothesis requires an alternative verification.
One can compare our spectral equations with the spectral equations for the lattice Bose gas, see \cite{S:BA}. The evolution model here and the lattice Bose gas are rather different systems, but both are related to a spin lattice, and remarkably there is something common in their spectral equations.

Also, we would like to discuss briefly our equations (\ref{mysym},\ref{mysp}). Let us consider this system as a system of equations for two independent sets $\{X_j\}$ and $\{x_j\}$. One can try to exclude the variables $X_N,X_{N-1},\cdots,$ from the system using e.g. the resultants. As the result, one obtains a polynomial equation for a single $X_j$. The degree of this polynomial is $N!$. Thus, the system (\ref{mysym},\ref{mysp}) as the system of equations for $\{X_j\}$ has $N!$ solutions. For instance, equations (\ref{X1}) and (\ref{XXZ}) represent just two branches out of $N!$ ones. However, we expect that all $N!$ branches of the solution are relevant. Anyway, a thermodynamic limit of the system (\ref{mysym},\ref{mysp}) requires a separate study.
\\

\noindent
\textbf{Acknowledgements.} The author would like to thank V. Bazhanov, R. Kashaev and V. Mangazeev for valuable discussions.

\appendix

\section{System of equations resulting the spectral equations}\label{AA}

This is rather technical section. Here we collect all equations leading to our conjecture (\ref{mysp}), and in particular to our main conjecture (\ref{myconj}). 

Let 
\begin{equation}
u_1,u_2,\cdots, u_N\;,\quad N=K\times L
\end{equation}
be the set of our spectral parameters. As the preliminary step, we need to distribute them to the points $(n_k,m_\ell)$ on the lattice,
\begin{equation}\label{lat}
\begin{tikzpicture}[scale=0.75]
\draw [-stealth, thin] (0,-1) -- (0,4);
\draw [-stealth, thin] (1,-1) -- (1,4);
\draw [-stealth, thin] (3,-1) -- (3,4);
\draw [-stealth, thin] (-1,0) -- (4,0);
\draw [-stealth, thin] (-1,1) -- (4,1);
\draw [-stealth, thin] (-1,3) -- (4,3);
\node [below] at (0,-1) {\scriptsize $n_1$};
\node [below] at (1,-1) {\scriptsize $n_2$};
\node [below] at (2,-1) {\scriptsize $\cdots$};
\node [below] at (3,-1) {\scriptsize $n_K$};
\node [left] at (-1,0) {\scriptsize $m_1$};
\node [left] at (-1,1) {\scriptsize $m_2$};
\node [left] at (-1,2) {\scriptsize $\vdots$};
\node [left] at (-1,3) {\scriptsize $m_L$};
\end{tikzpicture}
\end{equation}
Here we assume $n_1<n_2<\cdots N_K$ and $m_1<m_2<\cdots<M_L$, and also we assume that all these numbers $n_k,m_\ell$ are in general position. The way how we distribute parameters $u_j$, we represent it as the matrix 
\begin{equation}\label{umat}
\hat{u}\;=\;\left(\begin{array}{cccc}
u_{1,L} & u_{2,L} & \cdots & u_{K,L} \\
\vdots & \vdots & & \vdots\\
u_{1,2} & u_{2,2} & \cdots & u_{K,2} \\
u_{1,1} & u_{2,1} & \cdots & u_{K,1}
\end{array}\right)\;.
\end{equation}
Note that the matrix notation is non-standard, it corresponds to the picture (\ref{lat}).
The eigenstate (\ref{BA}) is  
\begin{equation}\label{APsi}
\Psi\;=\;\sum_{\hat{u}} C(\hat{u}) : \prod_{k,\ell} A_{n_k,m_\ell}(u_{n_k,m_\ell}) :\;,
\end{equation}
where "the sum over $\hat{u}$" means that the sum is taken with respect to all $N!$ permutations of the spectral parameters $u_1,\cdots, u_N$ inside the matrix (\ref{umat}). 

Now we are ready to describe the first step. Let us select the most left-bottom point $(n_1,m_1)$ in (\ref{lat}), then consider the set of differences $n_k-n_1$ and $m_\ell-m_1$ and rewrite them in the increasing order:
\begin{equation}
\{ n_k-n_1,m_\ell-m_1\}\;=\;\{ \delta_1<\delta_2<\cdots <\delta_{K+L-2}\}\;,
\end{equation}
and establish the correspondence 
\begin{equation}
\delta_1\to u_{\delta_1}\;,\;\; \delta_2\to u_{\delta_2}\;,\;\; \cdots\;.
\end{equation}
In more details, if $\delta_j=n_k-n_1$ for some $j$, then $u_{\delta_j}=u_{n_k,m_1}$, and if 
$\delta_j=m_\ell-m_1$ for some other $j$, then $u_{\delta_j}=u_{n_1,m_\ell}$. Let also for shortness
\begin{equation}
u_{\delta_0}\;=\;u_{n_1,m_1}\;.
\end{equation}

General formula for $A_{n_1,m_1}(\hat{u})$ is then (see (\ref{one-A}))
\begin{equation}
A_{n_1,m_1}(\hat{u}) \;=\; \aop_{2,(n_1,m_1)}^{+} + \sum_{\delta=1}^M g_{n_1,m_1}^{(\delta)}(\hat{u}) u_{n_1,m_1}^{-\delta} 
\bop_{n_1,m_1}^{(\delta)}\;,
\end{equation}
where for shortness
\begin{equation}
\bop_{n_1,m_1}^{(\delta)} \;=\; \aop_{1,(n_1+\delta,m_1)}^{+}\aop_{3,(n_1,m_1+\delta)}^{+}
\end{equation}
and where ``generic points'' of $A_{n_1,m_1}^{+}$ provide the conservation of the coefficients $g_{n_1,m_1}^{(\delta)}(\hat{u})$, similar to (\ref{one-sp}), but with the breaks:
\begin{equation}\label{chain}
\left\{\quad
\begin{array}{l}
\ds g_{n_1,m_1}^{(1)}(\hat{u}) =\cdots=g_{n_1,m_1}^{(\delta_1)}(\hat{u}) = \frac{1+qu_{\delta_0}}{1-q^2},\\
\\
\ds g_{n_1,m_1}^{(\delta_1+1)}(\hat{u})= \cdots = g_{n_1,m_1}^{(\delta_2)}(\hat{u})\;,\\
\\
\ds g_{n_1,m_1}^{(\delta_2+1)}(\hat{u})= \cdots = g_{n_1,m_1}^{(\delta_3)}(\hat{u})\;,\\
\\
\ds \vdots\\
\\
\ds g_{n_1,m_1}^{(\delta_{K+L-2}+1)}(\hat{u}) = \cdots = g_{n_1,m_1}^{(M)}(\hat{u}) = \frac{q+u_{\delta_0}}{1-q^2} u_{\delta_0}^M\;.
\end{array}\right.
\end{equation}
The staring and the ending points of this chain are fixed in the same way as in (\ref{one-sp}), while the breaks are described by relations 
\begin{equation}\label{2jump}
\begin{array}{l}
\ds g_{n_1,m_1}^{(\delta_{k+1})}(\hat{u}) \;=\; 
S(u_{\delta_k},u_{\delta_0}) g_{n_1,m_1}^{(\delta_k)}(\hat{u}) - \\
\\
\ds \qquad \qquad - \left(\frac{u_{\delta_k}}{u_{\delta_0}}\right)^{-\delta_k} 
\frac{C(\hat{u}'_k)}{C(\hat{u})} \frac{(q+u_{\delta_0})(1+qu_{\delta_0})}{q(u_{\delta_0}-u_{\delta_k})}
g_{n_1,m_1}^{(\delta_k)}(\hat{u}'_k)\;,
\end{array}
\end{equation}
where
\begin{equation}
S(u_{\delta_k},u_{\delta_0}) \;=\;
\frac{q+u_{\delta_k} + q^2 u_{\delta_k} + qu_{\delta_k}u_{\delta_0}}{q(u_{\delta_k}-u_{\delta_0})}\;\equiv\;
\frac{q^{-1}x_{\delta_k}-qx_{\delta_0}}{x_{\delta_k}-x_{\delta_0}}\;,
\end{equation}
and $\hat{u}'_k$ is the matrix $\hat{u}$ where the elements $u_{\delta_k}$ and $u_{\delta_0}$ are exchanged:
\begin{equation}
\hat{u}'_k\;=\;\hat{u} [u_{\delta_0} \leftrightarrow u_{\delta_k}]\;.
\end{equation}
Equation (\ref{2jump}) appears as the solution of 
\begin{equation}
\begin{array}{l}
\ds \boldsymbol{U} \left(
C(\hat{u}) g_{n_1,m_1}^{(\delta_k)}(\hat{u}) u_{\delta_0}^{-\delta_k} \bop_{n_1,m_1}^{(\delta_k)}
\left(\frac{q+u_{\delta_k}}{1-q^2} \bop_{\delta_k}^{(0)} + \aop_{2,\delta_k}^{+}\right) + 
[u_{\delta_0}\leftrightarrow u_{\delta_k}]\right)=\\
\\
\ds =\; u_{\delta_0}u_{\delta_k} \left(
C(\hat{u}) g_{n_1,m_1}^{(\delta_k+1)}(\hat{u}) u_{\delta_0}^{-\delta_k-1} \bop_{n_1,m_1}^{(\delta_k+1)}
\left(
\aop_{2,\delta_k}^+ + \frac{1+qu_{\delta_k}}{1-q^2} u_{\delta_k}^{-1} \bop_{\delta_k}^{(1)}\right)
+[u_{\delta_0}\leftrightarrow u_{\delta_k}]\right)\;,
\end{array}
\end{equation}
where what is inside big round brackets -- it is a proper part of the global eigenstate $\Psi$, and 
all other $A_{\pt}^{+}$-multipliers to it are in "general position".

The scheme (\ref{chain}) is a chain of relations (\ref{2jump}), they can be iterated, so that the starting and ending points of this chain are known. In this sense  the whole chain (\ref{chain}) gives \textbf{a single} linear equation for $C(\hat{u})$ for the given $\hat{u}$ and for several nearest $\hat{u}'_k$.

In the similar way one can define all sets $g_{n_k,m_\ell}(\hat{u})$ for the given fixed matrix $\hat{u}$ just by rewriting the lattice (\ref{lat}) by means of simple cyclic shifts, for example 
\begin{equation}\label{lat2}
\begin{tikzpicture}[scale=0.75]
\draw [-stealth, thin] (0,-1) -- (0,4);
\draw [-stealth, thin] (1,-1) -- (1,4);
\draw [-stealth, thin] (3,-1) -- (3,4);
\draw [-stealth, thin] (-1,0) -- (4,0);
\draw [-stealth, thin] (-1,1) -- (4,1);
\draw [-stealth, thin] (-1,3) -- (4,3);
\node [below] at (0,-1) {\scriptsize $n_1$};
\node [below] at (1,-1) {\scriptsize $n_2$};
\node [below] at (2,-1) {\scriptsize $\cdots$};
\node [below] at (3,-1) {\scriptsize $n_K$};
\node [left] at (-1,0) {\scriptsize $m_1$};
\node [left] at (-1,1) {\scriptsize $m_2$};
\node [left] at (-1,2) {\scriptsize $\vdots$};
\node [left] at (-1,3) {\scriptsize $m_L$};
\node [right] at (6,1.5) {$\to$};
\end{tikzpicture}
\qquad
\begin{tikzpicture}[scale=0.75]
\draw [-stealth, thin] (0,-1) -- (0,4);
\draw [-stealth, thin] (1,-1) -- (1,4);
\draw [-stealth, thin] (3,-1) -- (3,4);
\draw [-stealth, thin] (-1,0) -- (4,0);
\draw [-stealth, thin] (-1,1) -- (4,1);
\draw [-stealth, thin] (-1,3) -- (4,3);
\node [below] at (0,-1) {\scriptsize $n_2$};
\node [below] at (1,-1) {\scriptsize $n_3$};
\node [below] at (2,-1) {\scriptsize $\cdots$};
\node [below] at (3,-1) {\scriptsize $n_1+M$};
\node [left] at (-1,0) {\scriptsize $m_2$};
\node [left] at (-1,1) {\scriptsize $m_3$};
\node [left] at (-1,2) {\scriptsize $\vdots$};
\node [left] at (-1,3) {\scriptsize $m_1+M$};
\end{tikzpicture}
\end{equation}
As the result, one obtains $N\times N!$ linear equations (one equation for each vertex $(n_k,m_\ell)$ and for $N!$ permutations of $\hat{u}$). Hopefully, a rank of this system is essentially less, a lot of $C(\hat{u})$ remain free, however a part of self-consistency relations appear on this stage.
\\

The second step is related to the second type of linear equations. Nemely, let 
$(n_k,m_\ell)$ and $(n_{k'},m_{\ell'})$ be two points on the picture (\ref{lat}), such that $n_k<n_{k'}$ and $m_\ell<m_{\ell'}$ without loss of generality according to the principle (\ref{lat2}). Then the second type linear equation is 
\begin{equation}\label{2type}
\begin{array}{l}
\ds C(\hat{u}) g_{n_k,m_\ell}^{(\delta)}(\hat{u})g_{n_{k'},m_{\ell'}}^{(\delta')}(\hat{u}) u_{n_k,m_\ell}^{-\delta}
u_{n_{k'},m_{\ell'}}^{-\delta'} + \\
\\
\ds + C(\hat{u}') g_{n_k,m_\ell}^{(\delta)}(\hat{u}')g_{n_{k'},m_{\ell'}}^{(\delta')}(\hat{u}') u_{n_{k'},m_{\ell'}}^{-\delta}
u_{n_{k},m_{\ell}}^{-\delta'}\;=\;0\;,
\end{array}
\end{equation}
where $\hat{u}'$ is $\hat{u}$ with exchanged $u_{n_k,m_\ell}\leftrightarrow u_{n_{k'},m_{\ell'}}$, and where only two special values of 
$(\delta,\delta')$ are to be taken:
\begin{equation}
(\delta=m_{\ell'}-m_{\ell},\;
\delta'=n_k-n_{k'}+M)\quad \textrm{and}\quad
(\delta=n_{k'}-n_k,\;
\delta'=m_\ell-m_{\ell'}+M)\;.
\end{equation}
The origin of equation (\ref{2type}) is the following. If the right hand side of (\ref{2type}) is not zero, the Ansatz (\ref{APsi}) simply does not work.

The second step equations allow one to fix all remaining coefficients $C(\hat{u})$ and to produce all remaining self-consistency relations. 
Collecting finally all the self-consistency relations, one verifies that its system is equivalent to the system (\ref{mysym},\ref{mysp}), and the only result we could obtain is a particular form of polynomials $P_{n,N}(q)$.

We finalised this program for several cases including $N=2\times 2$ and $N=4\times 1$, when $4!=24$, what is a relatively small number. 
An attempt to verify our hypothesis for $N=3\times 3$ implies $9!=362880$, such number of unknowns for symbolical computations seems to be too big for a personal computer.

The final things to be mentioned are the following. The combination (\ref{APsi}) implies some additional details. For instance, when the creation operators $\aop_{1}^{+},\aop_3^{+}$ from different $A_{\pt}^+$ come to one point, the coefficients require more modification. Hopefully, such modifications do not affect the spectral equations.


\begin{thebibliography}{99}


\bibitem{Zam:80} 
A. B. Zamolodchikov, 
\emph{``Tetrahedra equations and integrable systems in three-dimensional space''},
Journal of Experimental and Theoretical Physics \textbf{52} (1980) 325


\bibitem{BS:82} 
V. Bazhanov and Yu. Stroganov, \emph{``On commutativity conditions for transfer matrices on multidimensional lattice''},
Theor. Math. Phys. \textbf{52} (1982) 685-691


\bibitem{BB:92} 
V. Bazhanov and R. Baxter,
\emph{``New solvable lattice models in three-dimensions''},
J. Stat. Phys. \textbf{71} (1993) 839-864


\bibitem{Kor:93} 
I. Korepanov,
\emph{``Tetrahedral Zamolodchikov algebras corresponding to Baxter's $L$-operators''},
Commun. Math. Phys. \textbf{154} (1993) 85–97


\bibitem{Hie:94} 
J. Hietarinta, \emph{``Labelling schemes for tetrahedron equations and dualities between them''},
J. Phys. \textbf{A27} (1994) 5727-5748


\bibitem{SMS:96} 
S. Sergeev, V. Mangazeeev and Yu. Stroganov,
\emph{The vertex reformulation of the Bazhanov-Baxter model''},
J. Stat. Phys. \textbf{82} (1996) 31-50


\bibitem{KV:91}  
M. M. Karpanov and V. A. Voevodsky, \emph{``$2$-categories and Zamolodchikov tetrahedra equations''}, 
Algebraic groups and their generalisations: quantum and infinite-dimensional methods (University Park,
PA, 1991), Proc. Sympos. Pure Math., vol. 56, Amer. Math. Soc., Providence, RI, 1994, pp. 177-259.


\bibitem{BS:2006} 
V. V. Bazhanov and S. M. Sergeev, \emph{``Zamolodchikov's tetrahedron equation and hidden structure of quantum groups''}, J. Phys. A: Math. Gen. 39 (2006) 3295-3310


\bibitem{S:2009} 
S. Sergeev, \emph{Super-tetrahedra and super-algebras}, J. Math. Phys. 50 (2009)
083519. 


\bibitem{Kun:2024} 
R. Inoue, A. Kuniba and Y. Terashima,
\emph{``Tetrahedron equation and quantum cluster algebras''},
J. Phys. A: Math. Theor. \textbf{57} (2024) 085202


\bibitem{BKMS:2025}
V. Bazhanov, R. Kashaev, V. Mangazeev and S. Sergeev, \emph{Quantum Dilogarithms and New Integrable Lattice Models in Three Dimensions}, 
\texttt{arXiv:2512.23338}, 2025


\bibitem{Bax-arc}
R. J. Baxter, private archive notes

\bibitem{K:95} 
I. G. Korepanov, \emph{Algebraic integrable systems, 2+1 dimensional models on wholly discrete space-time, and inhomogeneous models on 2-dimensional statistical physics}, Adv. PhD. Thesis, arXiv:solv-int/9506003, 1995




\bibitem{Bax-book} 
R. J. Baxter, ``\emph{Exactly Solved Models in Statistical Mechanics}''. Academic, London, 1982.


























\bibitem{S:2006}
S. Sergeev, \emph{Quantum curve in q-oscillator model}, 
International Journal of Mathematics and Mathematical Sciences
Volume 2006, Article ID 92064, Pages 1–31

\bibitem{Kuniba:2025}
R. Inoue, A. Kuniba, Y. Terashima, and J. Yagi,
\emph{Quantized six-vertex model on a torus},
\texttt{arXiv:2505.08924}

\bibitem{S:BA}
S. Sergeev, \emph{Ansatz of Hans Bethe for a two-dimensional lattice Bose gas},
J. Phys. A. \textbf{39} (2006) no 12 pp 3035-3045

\end{thebibliography}
\end{document}